\documentclass[12pt,preprint]{aastex}
\usepackage{natbib}
\usepackage{graphicx}

\shorttitle{2D evolution of self-gravitating magnetized tori}
\shortauthors{Fromang et al.}

\newcommand\bb[1] {   \mbox{\boldmath{$#1$}}  }
\newcommand\del{\bb{\nabla}}
\newcommand\bcdot{\bb{\cdot}}
\newcommand\btimes{\bb{\times}}

\begin{document}

\title{Evolution of self-gravitating magnetized disks. I- Axisymmetric 
        simulations}

\author{S\'ebastien Fromang \altaffilmark{1}}
\affil{Institut d'Astrophysique de Paris, 98Bis Bd Arago, 75014 Paris,
France}

\and

\author{Steven A.\ Balbus \altaffilmark{2} and Jean-Pierre De Villiers}
\affil{Virginia Institute of Theoretical Astronomys, Department of
Astronomy, University of Virginia, Charlottesville, VA 22903-0818}

\altaffiltext{1}{E-mail:fromang@iap.fr}
\altaffiltext{2}{Laboratoire de Radioastronomie, \'Ecole Normale
Sup\'erieure, 
24 rue Lhomond, 75231 Paris CEDEX 05, France}

\begin{abstract}
In this paper and a companion work, we report on the first global
numerical simulations of self-gravitating magnetized tori, subject
in particular to the influence of the magnetorotational instability
(MRI). In this work, paper I, we restrict our calculations to the study of
the axisymmetric evolution of such tori.  Our goals are twofold: (1)
to investigate how self-gravity influences the global structure
and evolution of the disks; and (2) to determine whether turbulent
density inhomogeneities can be enhanced by self-gravity in this regime.

As in non self-gravitating models, the linear growth of the MRI
is followed by a turbulent phase during which angular momentum is
transported outward.  As a result, self-gravitating tori quickly develop
a dual structure composed of an inner thin Keplerian disk fed by a
thicker self-gravitating disk, whose rotation profile is close to a
Mestel disk. Our results show that the effects of self-gravity enhance
density fluctuations much less than they smooth the disk, and giving
it more coherence.  We discuss the expected changes that will occur in 3D
simulations, the results of which are presented in a companion paper.
\end{abstract}


\section{Introduction}

\noindent
Accretion disks are the natural outcome of collapsing rotating
structures.   In some cases, their masses can be quite large and
self-gravitating effects may be important.   For example, observations
of water maser emission in NGC 1068 seem to suggest the presence of a
self-gravitating disk \citep{hure02,lodato03}.   On smaller spatial
scales, self-gravity is crucial in the final stages of star
formation.  Low angular momentum material is thought to collapse to a
protostar on a timescale of $10^5$ years, with higher angular
momentum
material forming a surrounding  disk \citep{Cassen81}.  Since the mass
of this disk grows as infall from the parent cloud continues, the disk
itself could, in principle, become self-gravitating.

The evolution of self-gravitating disks is strongly dependent upon the
value of the 
Toomre $Q$ parameter:

\begin{equation}
Q=\frac{c_s \kappa}{\pi G \Sigma} \, ,
\end{equation} 

\noindent
where $c_s$ is the sound speed, $\kappa$ the epicyclic
frequency and $\Sigma$ the disk surface density.  When $Q$
ranges between $1$ and $2$, numerical experiments show that
nonaxisymmetric instabilities, in the form of spiral density
waves, redistribute matter and transport angular momentum outwards
\citep{tohline&hachisu90,laughlin97,pickett00a,pickett03,mayer02}.

These calculations are purely hydrodynamical, and ignore magnetic
fields.
But the latter can be of great importance in a rotating gas, even when
the
fields are comparatively weak.  This is a result of the
magnetorotational
instability (MRI) \citep{balbus&hawley91,balbus&hawley98}, the outcome
of
which is a greatly enhanced outward turbulent angular momentum
transport.

We report here on the first MHD global simulations of a self-gravitating 
disks. We are interested in how the internal structure of the disk
evolves. 
Angular momentum transport is probably the most important process
affecting the disk, determining its size, surface density profile,
and surface emissivity.  There is, moreover, the possibility that
planets may form in the disk by gravitational collapse, as some recent
simulations suggest
\citep{boss97,boss98,mayer02}, though this notion is not entirely free
of
controversy \citep{tohline&hachisu90,pickett00b}.  The presence of MHD
turbulence
would certainly influence this issue.

As a first step, we begin our study of magnetized self-gravitating
disks by carrying out a series of axisymmetric numerical simulations
performed with the Zeus-2D code.  This restriction in dimensionality
precludes the growth of nonaxisymmetric gravitational instabilities,
and allows the initial focus to be centered on the MRI.  The obvious
shortcoming of this approach is that Cowling's Theorem prevents the
long term maintenance of MHD turbulence in 2D systems.  Experience with
global MHD codes has shown, however, that there is an extended period
of evolutionary development before substantial field decay occurs, and
that many of the flow features observed during this time are robust,
reappearing in fully 3D calculations.  Another important advantage is of
no small practical interest: in 2D it is possible to follow the global
evolution of these magnetized, self-gravitating systems with standard,
``in-house'' computational resources.  More computationally expensive
3D simulations are presented in a companion work, paper II of this series.

The paper is organized as follows.  In \S $2$, we describe our
initial equilibrium state and the numerical methods used.  Section 3
is a description of our results, including a comparison with non
self-gravitating calculations.  In section $4$, we summarize our
conclusions.

\section{Preliminaries}

\subsection{The initial configuration}
\label{toto}
\label{initial model}

The initial equilibrium for a self-gravitating disk satisfies the
equation of hydrostatic
balance, which reads, in standard cylindrical coordinates
$(\bb{e_r},\bb{e_{\phi}},\bb{e_z})$:

\begin{equation}
- \del P - \rho \del (\Phi_s+\Phi_c) + \rho r \Omega ^2 \bb{e_r} = 0 \, .
\end{equation}

\noindent
Here $\rho$ is the density, $P$ the pressure, $\Phi_s$ the
self-gravitating potential, $\Phi_c$ the potential created by 
a central mass $M_c$ and $\Omega$ the angular velocity.  Finding
an equilibrium
is not straightforward because of the global character of self-gravity;
an interative
method has proven to be relatively efficient.
This is the Self-Consistent Field (SCF) method developed
by \citet{hachisu86}.  The equation of state for the gas
is polytropic, and the angular velocity profile is fixed {\it a
priori}.  
We assume that the initial structure is a torus with
pressure and angular velocity profiles given by

\begin{eqnarray}
P & = & K_p \rho^{\gamma} \, , \\
\Omega & = & \Omega_0  (r/r_0)^{-q} \, ,
\end{eqnarray}

\noindent
where $K_p$ and $\Omega_0$ are constants. We chose an adiabatic
equation of state ($\gamma=5/3$), and took $q=1.68$, a value that
allows for a combination of self-gravity and pressure support.
The equilibrium boundary conditions are

\begin{equation}
\rho(R_{in})=\rho(R_{out})=0 \, ,
\end{equation}

\noindent
where $R_{in}$ and $R_{out}$ are respectively the inner and outer radii
of the torus.  For all models, we chose $R_{out}=1$, normalized the
density such that $\rho_{max}=1$, and took $G=1$. The SCF method then
involves guessing an initial density profile, and solving iteratively
for the density corrections and converged potential. The values of $K_p$
and $\Omega_0$ emerge in the process. 

\begin{figure*}
\plotone{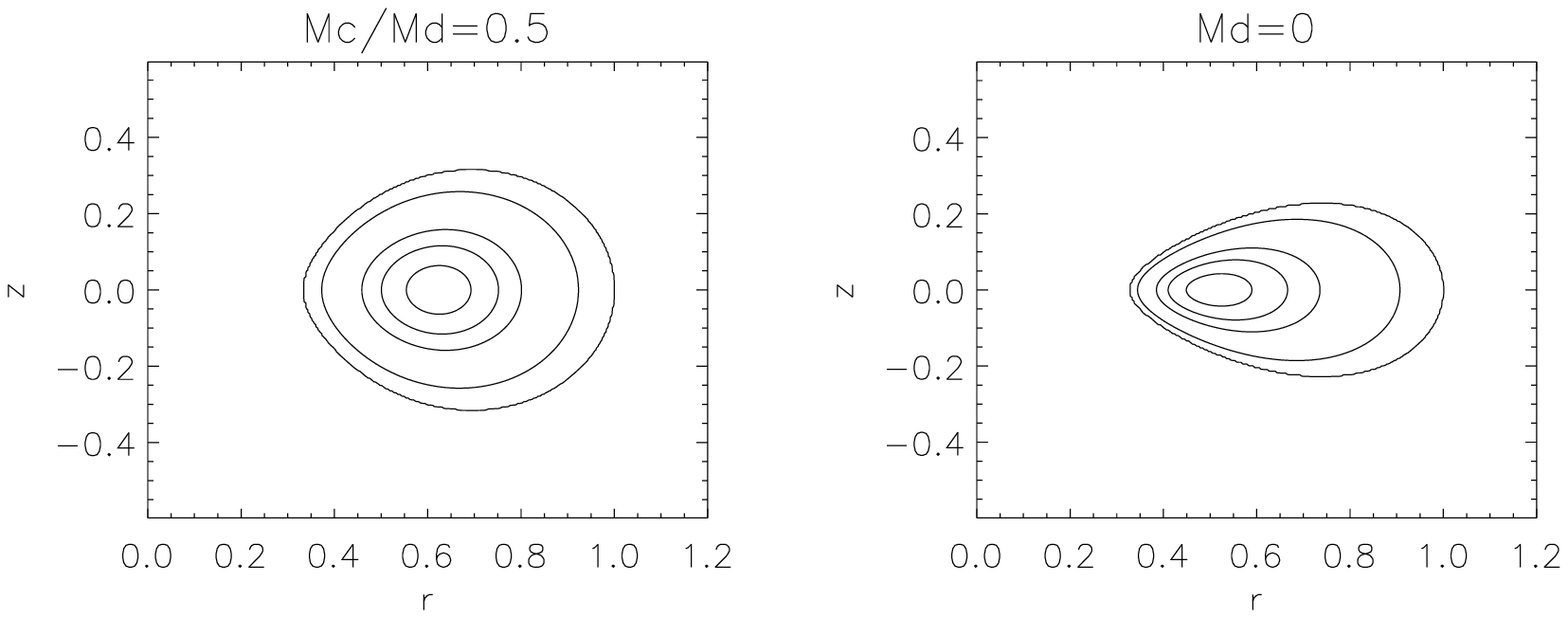}
\caption{Density contour in the (r-z) plane of 2 tori whose density
field has been computed using the SCF method.  The maximum density is
$\rho_{max}=1$ and there are 5 contour levels with
$\rho=10^{-4},0.1,0.5,0.7$ and $0.9$.  The calculations are done with $P
\propto \rho^{\gamma}$ with $\gamma=5/3$ and $\Omega \propto r^{-q}$
with $q=1.68$.  The left hand side model has a central mass whose mass
is
half that of the torus, and the right hand side model is a zero mass
torus.}
\label{tori structure}
\end{figure*}

In Figure \ref{tori structure}, we compare the result of such a
procedure with the density field of a zero mass torus.  Both models
have $R_{in}=1/3$.  The left panel shows the density contours of the
self-gravitating disc.  A central point mass containing half the mass of
the torus is assumed to be present.  The right panel shows the density
contours of the massless disc.  The gravitational potential in this case
is due
only to a central point mass, which was taken to be $GM_c=0.35$
, chosen so that
the angular momentum radial profiles of both models were very close.
This is
important if we want to have similar MRI growth rates. Although the
``massless''
model is more elongated because of the central concentration of the
Keplerian potential, the density field of both models are rather
comparable.
They will be used as starting points of the following calculations.

At the beginning of the numerical simulations, a weak poloidal magnetic
field, confined to lie inside the
torus, is added to the equilibrium structure.
We follow the method described in \citet{hawley00}, with 
the toroidal component of the potential vector given by:

\begin{equation}
A_{\phi} \propto \rho \cos \left( 2\pi 
\frac{r-R_{in}}{R_{out}-R_{in}} \right) \, .
\label{Aphi}
\end{equation}
Radial and vertical component of the magnetic field are then normalized
so
that the volume averaged value of the ratio of gas pressure to magnetic
energy $\langle \beta \rangle$ equals a predetermined value.  
 The new magnetized configuration is no longer in equilibrium, and the
resulting small disturbances are sufficient to trigger the MRI.

\subsection{Algorithms}

The equations of ideal MHD are:

\begin{eqnarray}
\frac{\partial \rho}{\partial t} + \del \bcdot (\rho \bb{v})=0 \, ,\\
\rho \left( \frac{\partial \bb{v}}{\partial t} + \bb{v} \bcdot \del
\bb{v} \right) = - {\del} P - \rho {\del} \Phi + \frac{1}{4
\pi} (\del \btimes \bb{B}) \btimes \bb{B} \, ,\\
\rho \left( \frac{\partial }{\partial t} + \bb{v} \bcdot \del \right)
\left( \frac{e}{\rho} \right) = -P \del \bcdot \bb{v} \, ,\\
\frac{\partial \bb{B}}{\partial t} = \del \btimes ( \bb{v} \times
\bb{B} ) \, .
\label{MHD equations}
\end{eqnarray}

\noindent
Here $\rho$ is the density, $e$ the energy density, $\bb{v}$ the
fluid velocity, $\bb{B}$ the magnetic field, $P$ the pressure and
$\Phi$ the total gravitational potential, i.e. the sum of the
self-gravitating and the central mass potentials. We have carried out
our numerical simulations with the Zeus-2D code.

Zeus-2D has been described in detail by
\citet{stone&norman92a,stone&norman92b}.  It solves the above equations
using time-explicit Eulerian finite differences.  Different geometries
are allowed through the use of a covariant formalism; here we use
cylindrical coordinates throughout.  The magnetic field is evolved using
the constrained transport method \citep{evans&hawley88}, which
guarantees
the divergence-free constraint to be satisfied at all times if it is
satisfied initially.  Electromotive forces are calculated using the
method of characteristics so that Alfv\'en wave propagation is
accurately
calculated \citep{stone&norman92b}. Different boundary conditions are
implemented. In the simulations reported here, we have used outflow
boundary conditions everywhere.

We have used the code in its original form, except for the calculation
of $\Phi_s$.  We detail here the procedure used. 

To calculate the gravitational potential due to an isolated distribution
of matter, one proceeds in two steps.  First, $\Phi_s$ is calculated on
the boundary of the computational domain by a direct Green's function
expansion (see below), and then it is calculated on the whole grid by
the Succesive Over Relaxation Method (SOR) described in \cite{hirsh88}.
While rapid algorithms have been developped for the second step, the
first remains very time consuming.  Traditionally, spherical harmonics
have been used as the basis for the Green's function expansion
\citep{stone&norman92a,boss&myhill95,muller&steinmetz95,yorke&kaisig95}
even when the simulations are not performed in spherical geometry.
Recently, \citet{cohl&tohline99} have argued that a Legendre function
basis is better suited to cylindrical geometry.  They give a very
compact formula for the gravitational potential of an axisymmetric
matter
distribution, well suited for axisymmetric numerical simulations in the
$(r,z)$ plane:

\begin{equation}
\Phi_s (r,z) = - \frac{2G}{\sqrt{r}} \int dr'dz' \sqrt{r'} \rho (r',z')
\mu K(\mu ) \, ,
\label{cohl and tohline}
\end{equation}

\noindent
where the integral has to be taken over the whole computational domain
(r,z).  Here

\begin{equation}
\mu = \frac{r^2+r'^2+(z-z')^2}{2rr'}
\end{equation}

\noindent
and $K$ represents the complete elliptic integral of the first kind
\citep{mathfunc}.  We have used equation \ref{cohl and tohline} in our
simulations to calculate $\Phi_s$ on the boundary.  Inside the
computational domain, we then solved the Poisson equation using the
SOR method cited above. 

Even with the advantages of this procedure, the computation of the
gravitational potential is still very expensive.  For high resolution
runs, we re-evaluated the potential only when it had changed by more
than a threshold value \citep{stone&norman92a}.  Comparison with lower
resolution runs in which the potential was updated at every time step
suggests that the overall results are not significantly affected by
this approximation.

\subsection{Diagnostics}

Here we define the vertical and volume averages used to
analyse the calculations.   The simplest is
$\langle \beta \rangle$, mentioned in section \ref{toto} (remember that 
the symbol $\langle . \rangle$ stands for a volume average):

\begin{equation}
\langle \beta \rangle = \frac{\langle P \rangle}{\langle B^2/8\pi
\rangle} \, .
\end{equation}

To make a connection with the standard disc theory, we use height
averages
of the Maxwell and Reynolds stresses, which will be denoted using an
overbar.
The Maxwell and Reynolds stresses will respectively be
calculated using \citep{hawley00}:

\begin{eqnarray}
T^{Max}_{r\phi}(r,t) & = & - \frac{\overline{B_rB_{\phi}}}{4\pi} \, ,\\
T^{Rey}_{r\phi}(r,t) & = & \overline{\rho v_r
v_{\phi}}-\frac{\overline{\rho v_r} \textrm{ } \overline{\rho
v_{\phi}}}{\overline{\rho}} \, .
\end{eqnarray}

\noindent
Note that 
in 3D, self-gravity would also produce a stress that 
gives rise to angular momentum transport, but not in axisymmetry.
When the Maxwell and Reynolds stress tensors
are normalized by
the pressure, the standard $\alpha$ parameter 
\citep{shakura&sunyaev73} emerges:

\begin{equation}
\alpha(r,t)=\frac{T^{Max}_{r\phi}+T^{Rey}_{r\phi}}{\overline{P} } \, .
\end{equation}

\noindent
A volume-averaged Maxwell stress tensor can also be defined:
\begin{equation}
\langle T^{Max}_{r\phi} \rangle (t) = \frac{\int_{R_{in}}^{R_{out}}
T^{Max}_{r\phi}(r,t) r dr}{\int_{R_{in}}^{R_{out}} rdr} \, ,
\end{equation}
with similar definitions for the volume-averaged Reynolds stress and
for $\langle \alpha \rangle$.

\section{Results}
\subsection{Model description}

The parameters of the simulations we performed are summarized in table
\ref{models}.  Model {\it{T}} is a hydrodynamical run done as a control
study.  No axisymmetric Jeans instability should be present, and none
was found.
(The integration time was 10 orbits at the initial location of the
pressure maximum.)

In models A2a and A2b, our fiducial runs, a central mass with half
the mass of the disc is present.  Initially, the angular velocity has
a non-Keplerian power law profile, with $\Omega \propto r^{-1.68}$.
The magnetic field is initialized as described above, with an initial
$\langle \beta \rangle$ of 1500.  The grid resolution is $128\times 128$
for model A2a and $256\times256$ for model A2b. Both
models give the same qualitative results; we focus now on the high
resolution run.

\begin{figure}
\begin{center}
\plotone{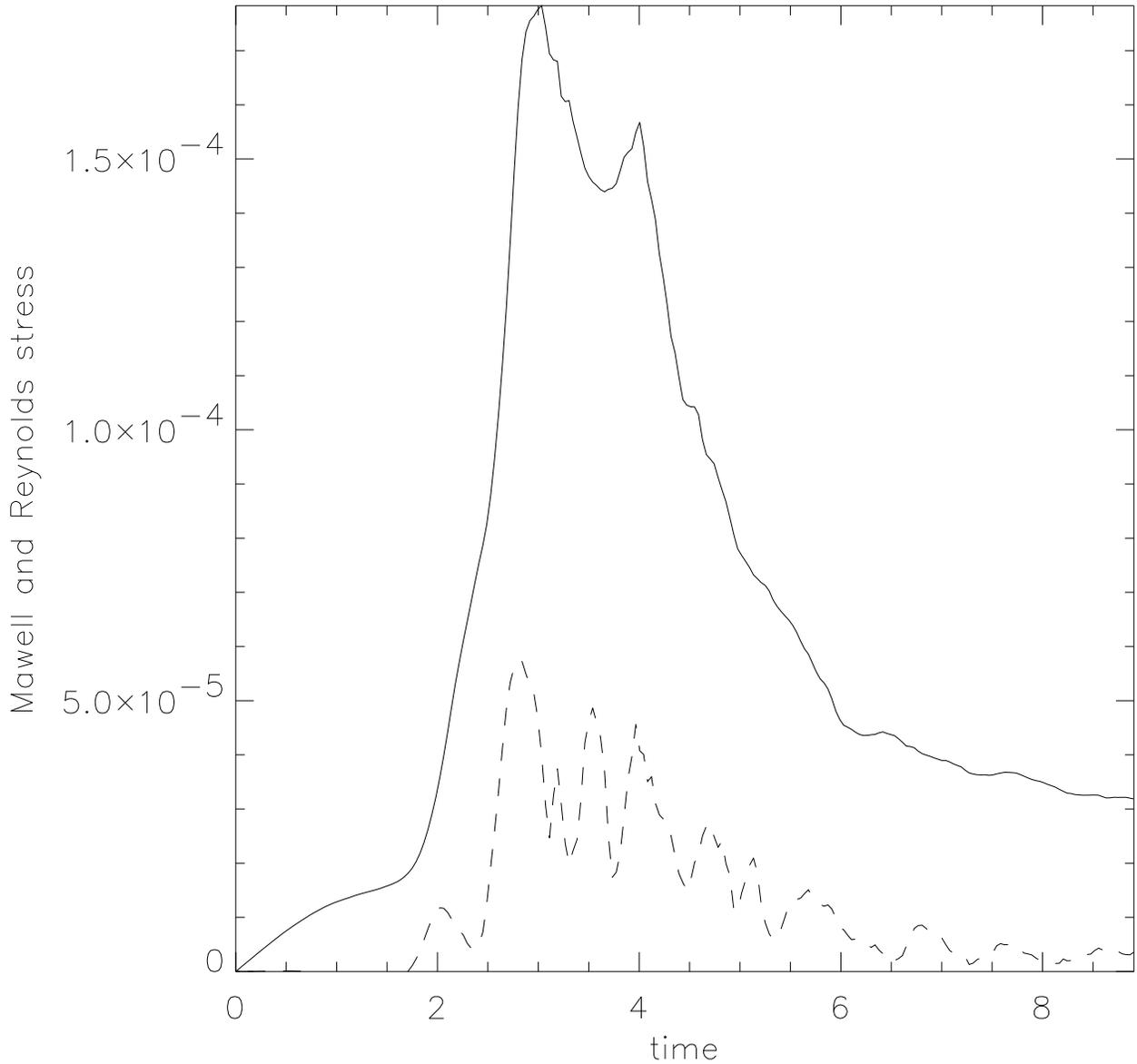}
\caption{Maxwell ({\it solid line}) and Reynolds ({\it dashed line})
stress  history in model A2b.  The time is measured in units of the
initial orbital time at pressure maximum.  The initial growth of the MRI
is shown by the rapid buildup of the Maxwell stress, and is followed by
a turbulent phase during which the Maxwell stress decays because of the
anti-dynamo theorem.  In agreement with simulations of non
self-gravitating discs, the Reynolds stress tensor is always 
smaller than the Maxwell stress.}
\label{run history}
\end{center}
\end{figure}

Figure \ref{run history} shows the volume averaged Maxwell ({\it solid
line}) and Reynolds ({\it dashed line}) stress tensor history for model
A2b.  The Maxwell stress shows the initial growth typical of the MRI. 
One
can see during this phase the development of the radial streaming
structures often observed in local and global simulations of zero mass
discs \citep{hawley00}.  The linear MRI saturates after $3$ orbits, and
then breaks down into turbulence.  In
accord with the non self-gravitating results, figure \ref{run
history} also shows that the Reynolds stress has a significantly smaller
amplitude than the Maxwell stress at all times: most of the angular
momentum transport is due to the magnetic stress.

\begin{figure}
\begin{center}
\plottwo{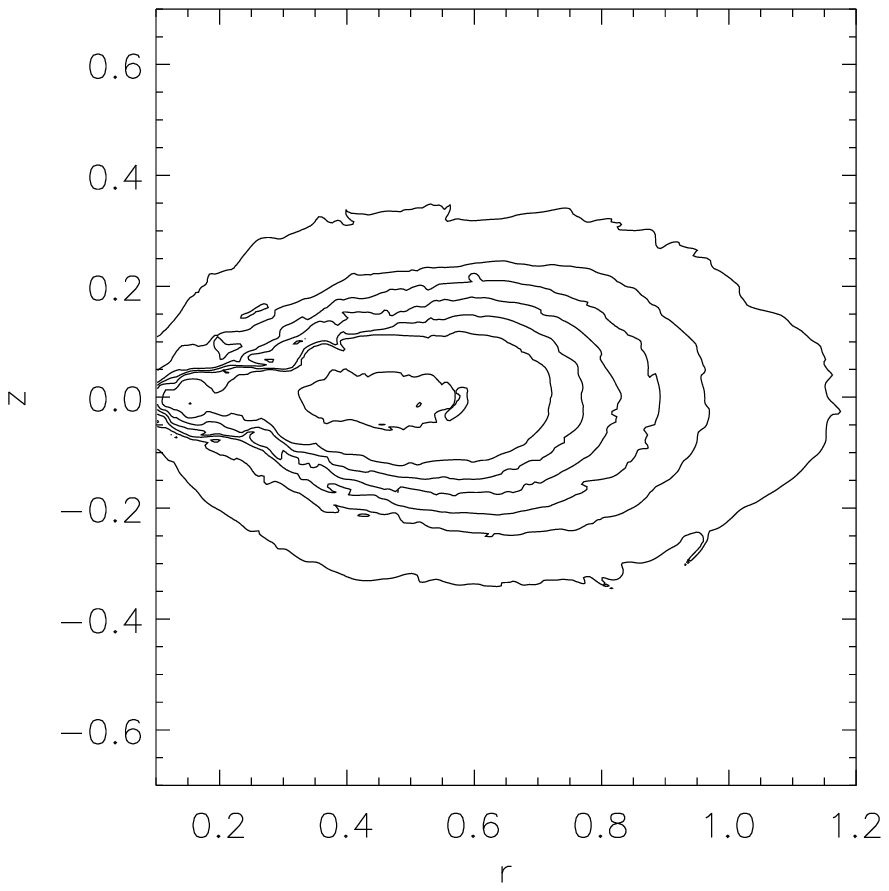}{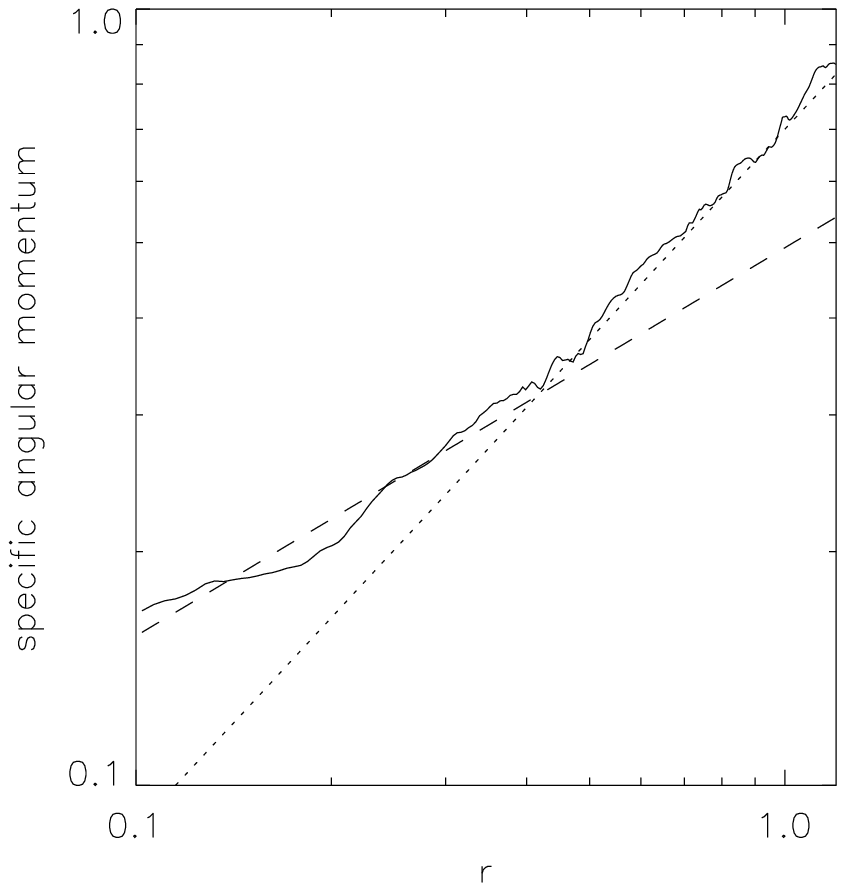}
\caption{The left hand side panel shows the density contour in the (r-z)
plane for model A2b after 5.85 orbits, normalized by the maximum
density.  There are 7
levels: $\rho / \rho_{max}=0.01,0.1,0.2,0.3,0.4,0.5,0.7$.  The right
panel shows
the specific angular momentum profile ({\it solid line}) for the same
model averaged between orbits $5.05$ and $7$.  The dashed line shows the
Keplerian profile expected for the central mass, while the dotted line
is
a fit of the outer part of the disc, with $l \propto r^{0.9}$.  }
\label{ang_mom profile}
\end{center}
\end{figure}

\begin{figure*}
\begin{center}
\plottwo{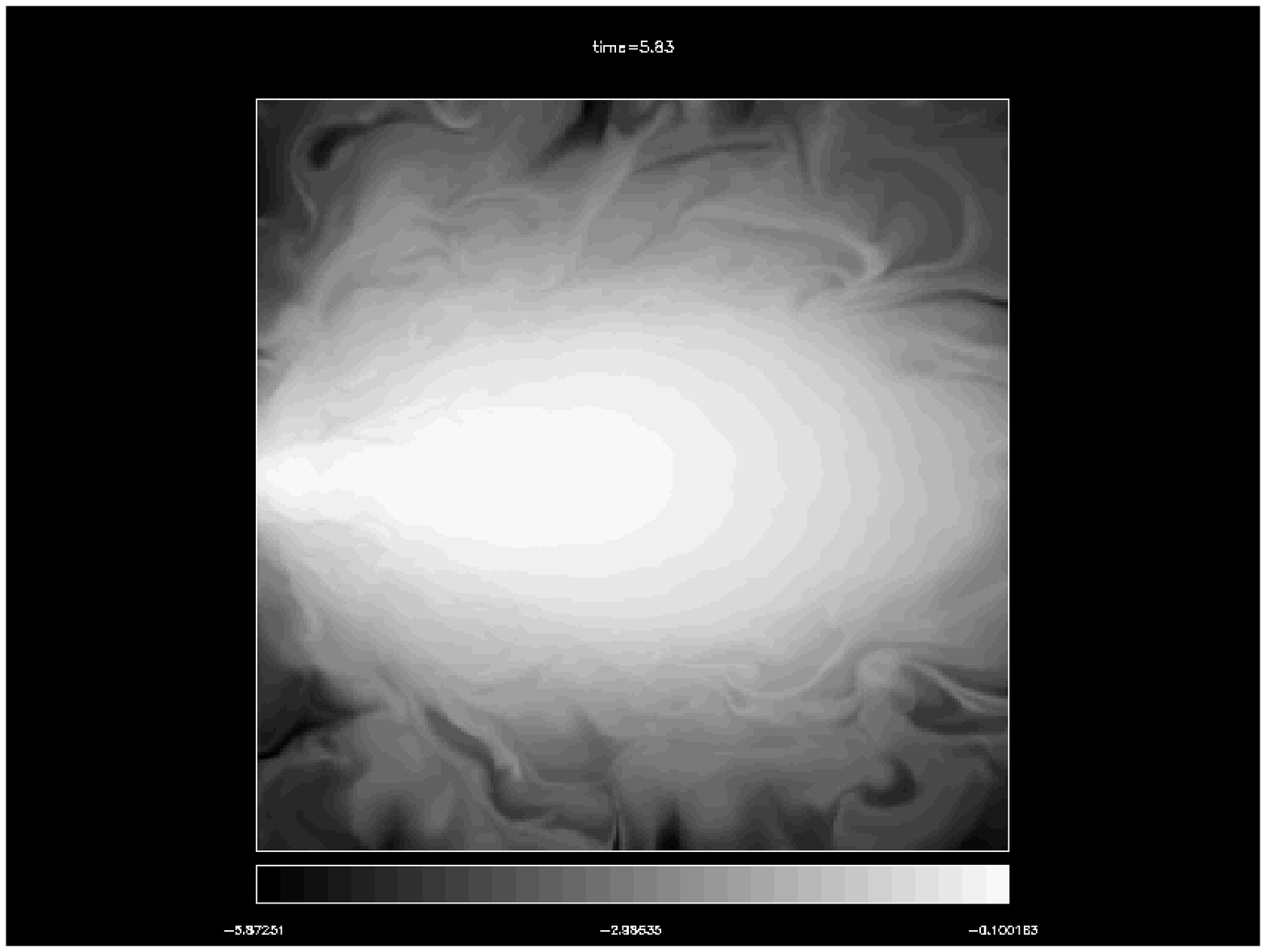}{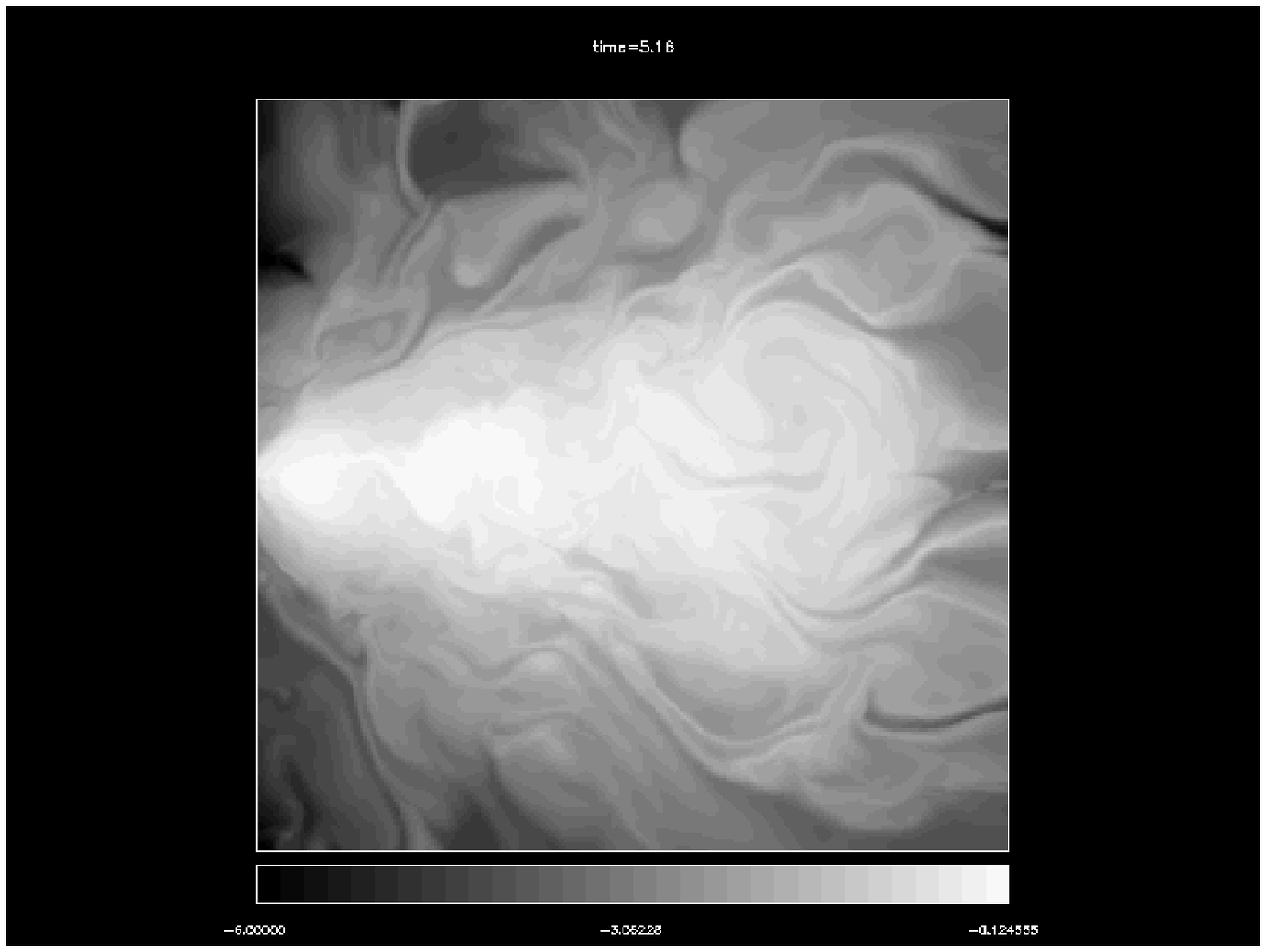}
\caption{Snapshots of the density logarithm in the self-gravitating
model A2b ({\it left
panel}) after 5.85 orbits at the initial pressure maximum and in the non
self-gravitating model
B ({\it right panel}) after 5.94 orbits.  The self-gravitating torus has
developed an inner Keplerian thin disc fed by an outer thick ring.  The
zero mass model shows a much less coherent structure, with large density
fluctuations, and an approximately constant $H/r$ value.}
\label{model A2 snapshots}
\end{center}
\end{figure*}

In Figure \ref{ang_mom profile} (left panel), we show the contours of
the
density distribution in the $(r-z)$ plane for model A2b at time $5.85$,
just after turbulence has set in (cf also the left hand side of
Figure \ref{model A2 snapshots}).  The initial torus has developed a
dual structure composed of an inner thin disc, through which matter is
accreted toward the central point mass, and an outer thick torus that
feeds the inner disc.  The gravitational potential is dominated by the
central star in the inner thin disc, and by self-gravity in the outer
thick disc.  Figure \ref{ang_mom profile} ({\it right panel}), shows
that
the rotation profiles are consistent with this.  We plot the angular
momentum profile in the equatorial plane of the disc averaged between
orbits $5.05$ and $7$ ({\it solid line}).  The inner disc is in
Keplerian
rotation around the central mass ({\it dashed line}) out to a radius of
$0.4$.
Beyond this point, a power law with exponent $0.9$ can be fit to the
profile ({\it dotted line}), close to the constant $V_{\phi}$ profile
of a fully self-gravitating Mestel disc
\citep{binney&tremaine87,bertin99}.

In our disk model, $Q_{min} \sim 0.3$, which means that the smallest 
unstable wavelength $\lambda_{min}$ is roughly equal to $0.5$, which is 
smaller than the radial extent of the disk. In spite of the presence of 
these unstable wavelengths, self-gravity does not seem to be able to enhance 
local MRI density inhomogeneities. 

\subsection{Comparison with zero mass tori}

Since they have very different structures, the question of how to draw
a comparison between self-gravitating and non-self-gravitating runs is
not completely straightforward.  For example, neither choosing similar
initial $\beta$ values nor saturated $\alpha$
values is satisfactory, since it greatly over emphasizes the (rather
minimal) role of gas pressure.  More revealing is the value of the
stress compared with the rotational energy it is attempting to disturb.
Accordingly, we define the ratio $R$ by:

\begin{equation}
R=\frac{\langle T^{Max}_{r\phi} \rangle + \langle T^{Rey}_{r\phi}
\rangle}{\langle \rho v_{\phi}^2 \rangle} \, .
\end{equation}

\begin{figure}
\begin{center}
\plotone{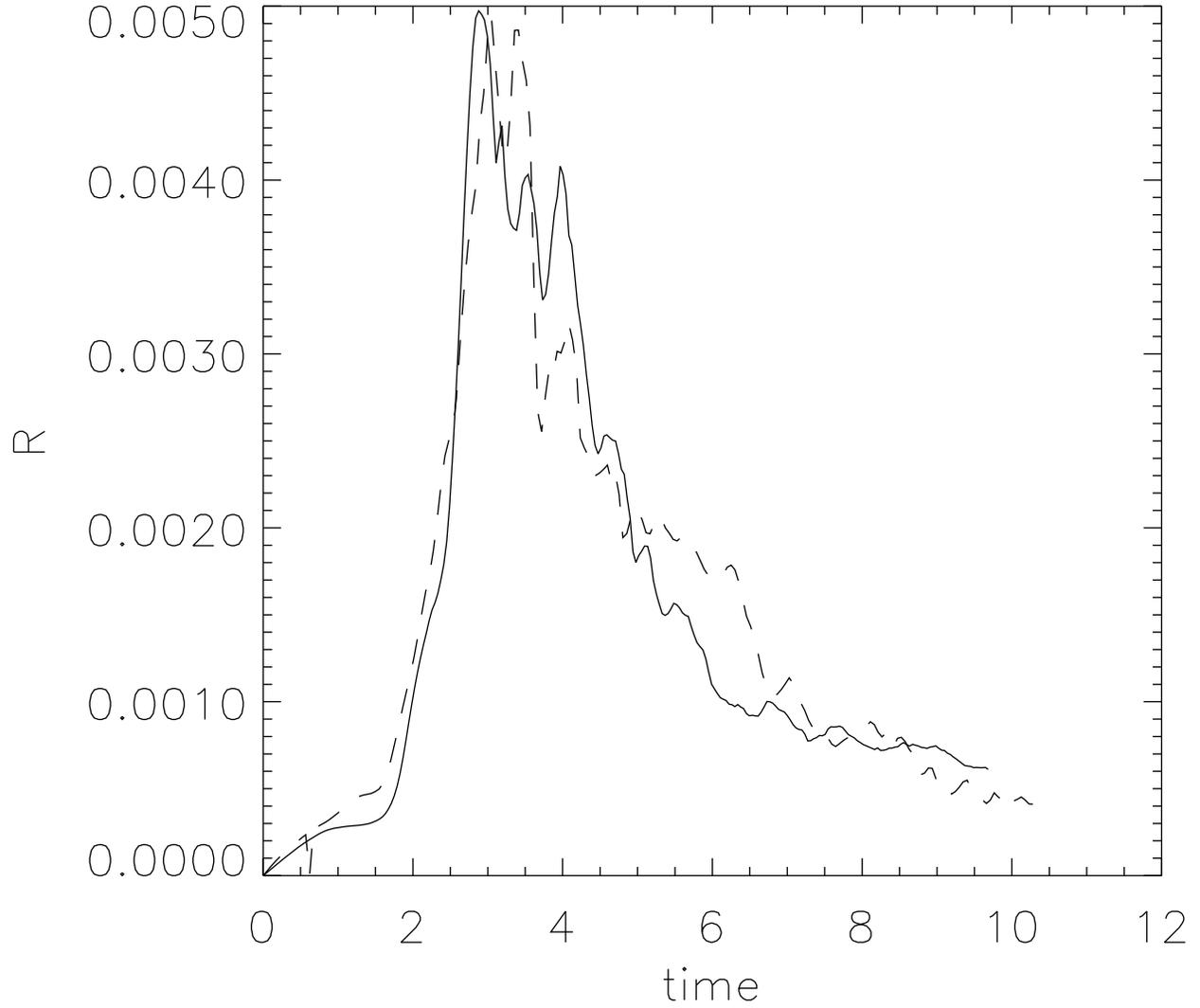}
\caption{Comparison between the time evolution of $R$ (See text) in
models A2b ({\it solid line}) and B ({\it dashed line}). The values
obtained are very close in both runs throughout the simulations.}
\label{compar_256_first.ps}
\end{center}
\end{figure}

\noindent
By design,  $\langle \rho v_{\phi}^2 \rangle$ are similar in each of
the initial models.  Consequently, the value of $R$ will be determined
by the level of angular momentum transport.  As noted, the MRI growth
rates will be nearly the same in the two models, because the angular
velocity profiles are similar.  We have found that an initial
$\langle\beta\rangle$ value of 200 in our zero mass model
gives a very similar evolutionary $R$ profile to our high resolution
model A2b (see fig.\ \ref{compar_256_first.ps}).  
We refer to this zero mass run as model B.

Density displays for model A2b after $5.85$ orbits ({\it left panel})
and
for model B after $5.94$ orbits ({\it right panel}) are shown in figure
\ref{model A2 snapshots}.  The scale is logarithmic.  Both cases are
fully
turbulent, but the difference is visually striking: model A2b clearly
has a much
more coherent internal structure than does model B. An FFT of the radial
fluctuations in the disc midplanes shows that on a scale of $\sim 1/15$
of the grid size, the non self-gravitating amplitudes are a factor of
$\sim 2$ larger than those of the self-gravitating run. This is comparable 
to what is seen in the lower resolution model A2a. It is possible that the
long range nature of Newtonian gravitational forces is responsible
for the higher degree of coherence seen
in self-gravitating discs compared with
zero mass discs. It should be noted, however, that
self-gravitating models also require a higher equilibrium pressure,
and this tends to enhance the size of coherent structures, as well 
\citep{sanoetal04}.

\subsection{Effect of the variation of the field strength}

Finally, we consider the effect of the initial field strength on the
evolution of the massive torus. We compare model A1 ($\langle \beta
\rangle=400$) and A2b ($\langle \beta \rangle=1500$).  Everything else
is kept the same.

\begin{figure}
\begin{center}
\plotone{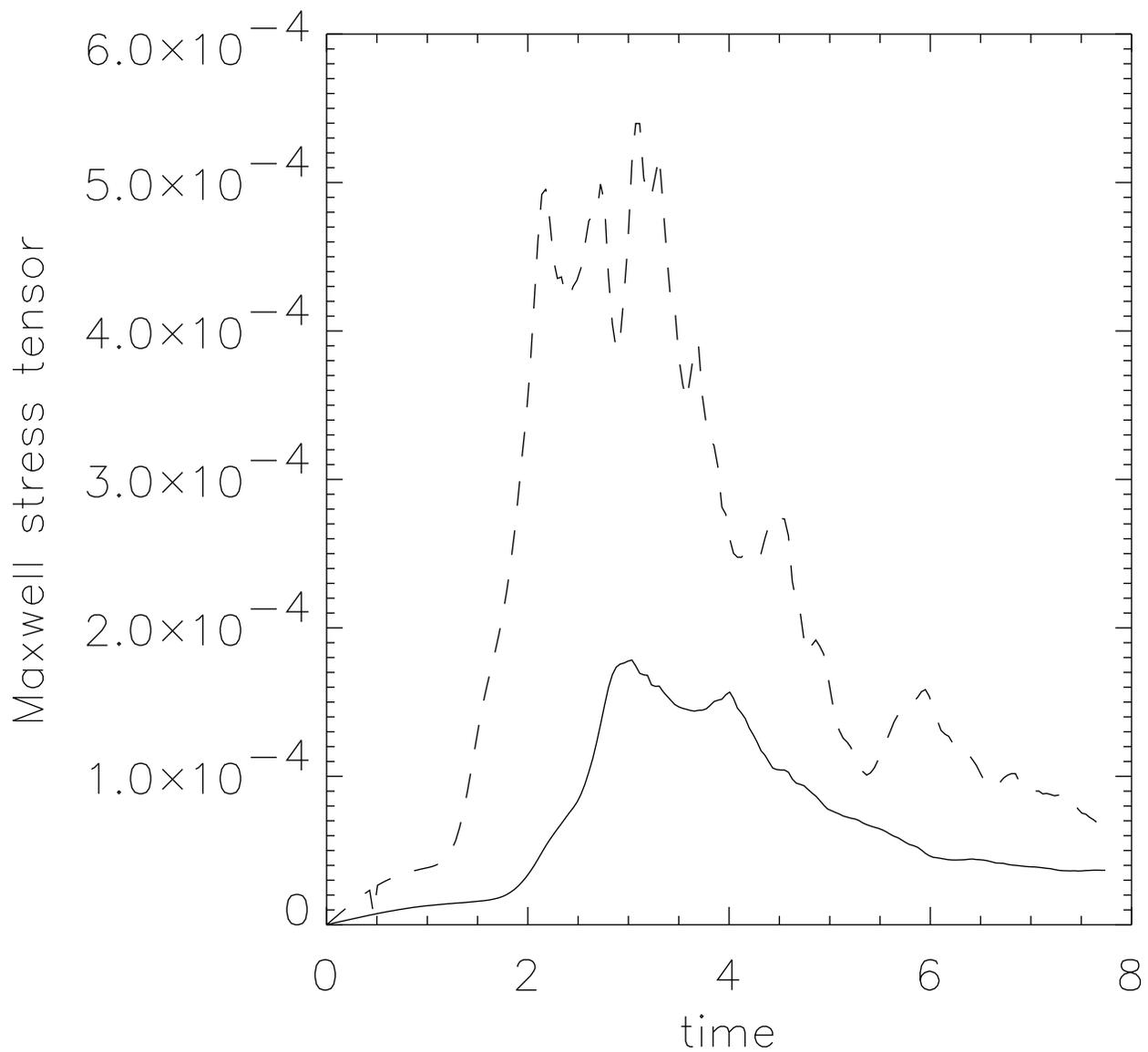}
\caption{Comparison between the time history of the Maxwell stress
tensor in model A2a ({\it solid line}) and in model A1 ({\it dashed
line}).}
\label{selfgrav compar}
\end{center}
\end{figure}

The Maxwell stress history is shown in figure \ref{selfgrav compar}.
$\langle T_{r\phi}^{Max} \rangle$ is larger in model A1 ({\it dashed
line})
during the whole simulation. It exceeds the value reached in model A2b
({\it solid line}) by a factor of about 3 at the maximum stress level.
A larger stress results in a larger mass accretion rate in model A1
compared with model A2b. This behaviour of retaining a memory of the
initial field strength is qualitatively similar to what has been found
before in 2D simulations of zero mass discs \citep{stone&pringle01}.

\begin{figure}
\begin{center}
\plottwo{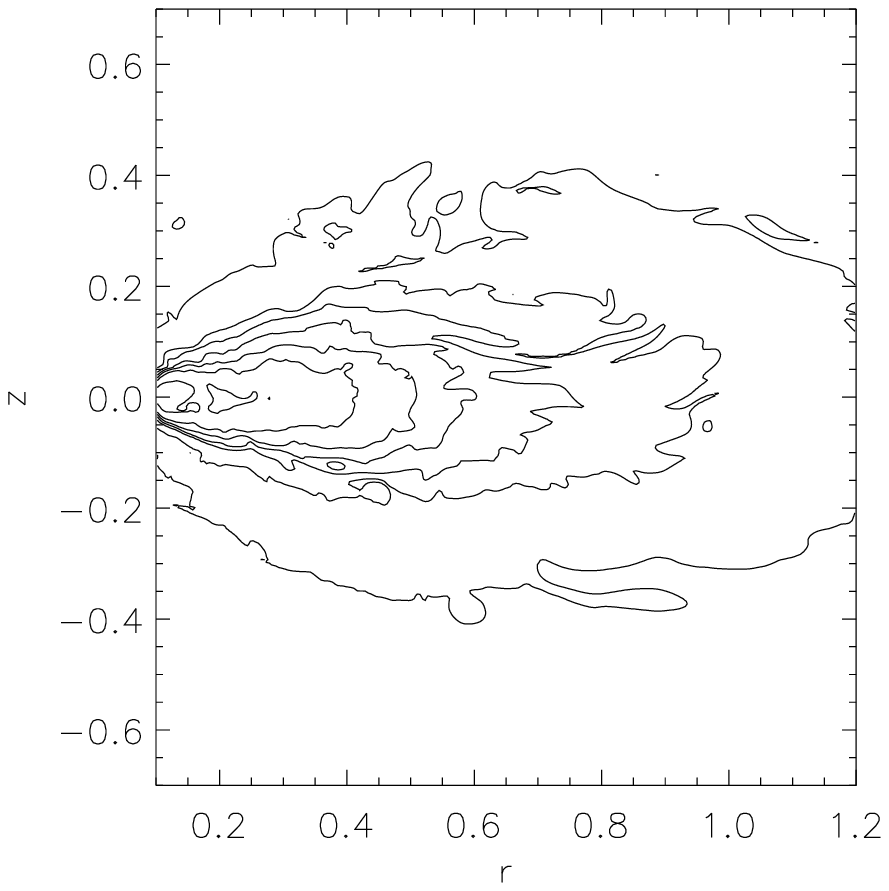}{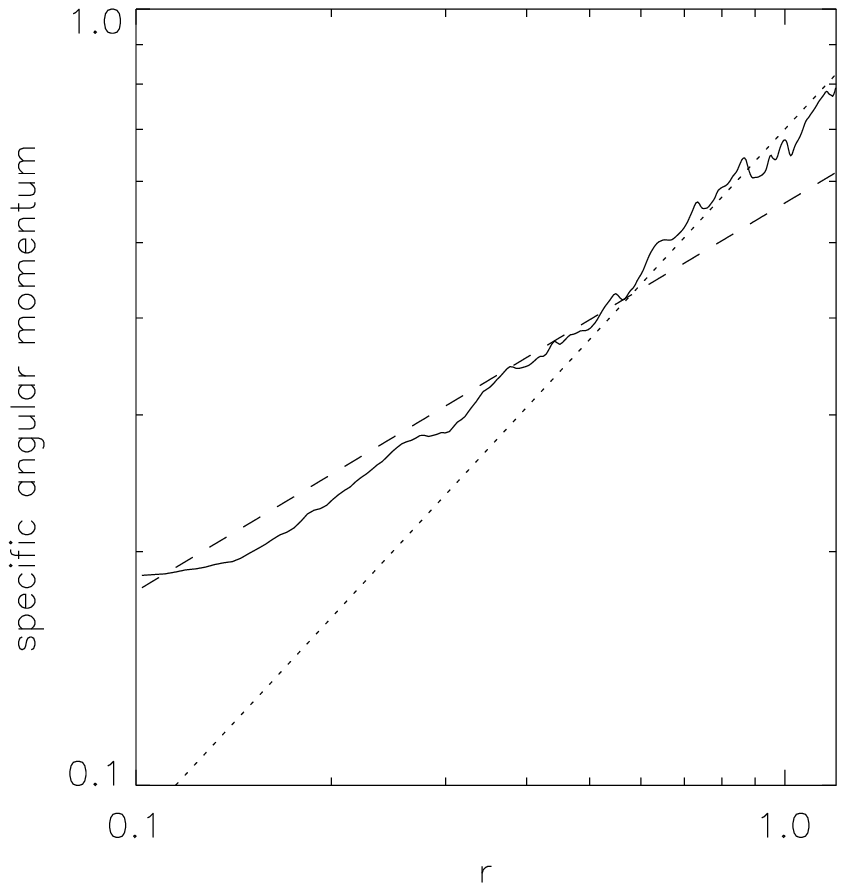}
\caption{Same as figure \ref{ang_mom profile}, but for model A1. The
left hand side plot is taken after 4.5 orbits at the initial maximum
pressure location and the angular momentum shown on the right hand side
was averaged between the orbits 3.88 and 5.05.}
\label{selfgrav disrupted}
\end{center}
\end{figure}

Figure \ref{selfgrav disrupted} shows the state of model A1 after $5.83$
orbits (it should be compared with model A2b in figure \ref{ang_mom
profile}). Density
contours are presented on the left side and the angular momentum
radial profile on the right side. The density structure at this
stage is more disturbed than in model A2b, showing a larger level
of MHD turbulence. However, the disc naturally adopts the same radial
distribution of angular momentum as in model A2b: Keplerian in the
inside, and ``Mestellian'' in the outer parts.  Note that the transition
radius is larger in model A1 than in model A2b. This is because the mass
accretion rate is larger in the former and the central point mass
consequently greater; self-gravity is less important in this case.

To conclude: while the simulations retain a memory of their initial
field strength and show different levels of turbulent stress, both tend
toward a final state of an inner Keplerian rotational profile and an
outer Mestel-like self-gravitating rotation profile.

\section{Discussion and Conclusion}

In this paper, we have investigated the qualitative behaviour of the
MRI in a self-gravitating tori.  We performed 2D numerical simulations
of the evolution of weakly magnetized, massive tori.  The simulations
are constructed so that no gravitational instability develops. As a
consequence, self-gravity cannot transport angular momentum outward as
in 3D.  The torus evolves only because of the effect of the MRI induced
MHD turbulence.  We found that the MRI behaves in these massive discs
qualitatively much like it does in the non self-gravitating disc, though
there are differences in the density response (see below).

Self-gravitating magnetized tori evolve toward a structure composed of two
parts: an inner thin disc in Keplerian rotation around a central mass,
fed by an outer more massive thick torus.  The gravitational potential
in the torus is dominated by the self-gravitating component of the
potential and it is no longer Keplerian.  Rather, its velocity profile
is approximately that of a Mestel disc, $v_{\phi}= {\rm constant}$. This
steepening of the specific angular momentum profile at large radii has
been seen in VLBI observations of water maser emission in the active
galactic nuclei NGC 1068 \citep{greenhill96}. In this case, the best fit
to the angular momentum profile is $j \propto r^{0.69}$.  Although we
don't find exactly the same radial profile, our result is quite close
(we obtain $j \propto r^{0.9}$). The disagreement probably comes from the
fact that the mass ratio between the central object and the disk in our
simulation and in the actual system are different. In the latter, the disk
mass is comparable to the central black hole mass \citep{lodato03}. In
our case, the mass of the disk is twice the mass of the central object,
thereby giving larger rotational velocities and a steeper profile.

In an attempt to highlight the difference between self-gravitating and
zero mass discs, we compared two such simulations with similar density
and rotation profiles, and similar evolutionary histories of their
stress tensors.  The appearance of the zero mass disc was, however,
considerably more disrupted than that of the self-gravitating disc.
The former showed large density fluctuations, while the latter maintained
a much more globally coherent structure. One might have expected to
see density fluctuations locally enhanced because of the presence of
self-gravity. On the contrary, this is the global nature of the potential
that seems to be more important in affecting the evolution of the disk:
in this regime, it smoothes the effect of the turbulence and gives more
coherence to the disk.

Of course, there are imporant limitations to this study imposed by
axisymmetry.  First, the turbulence is not sustainable because of the
anti-dynamo theorem: it decays and the two component Kepler-Mestel
structure described above evolves imperceptively after a few dynamical
timescales.  Follow-up 3D calculations are essential to track the
evolution of this two-component disc.  Even more importantly, 3D
calculations will allow the development of nonaxisymmetric structure,
and allow a full investigation of the interaction between MHD turbulence
and spiral structure gravitational instabilities.  This important problem
is the subject of the companion paper following this one.

\section*{ACKNOWLEDGMENTS}
We are grateful to Caroline Terquem for a critical reading of an earlier
version of this paper, and for constructive comments.
SF  acknowledges partial support from the Action Sp\'ecifique de
Physique Stellaire and the Programme National de Plan\'etologie and the
EC RTN Programme HPRN-CT-2002-00308. SAB is grateful to the Institut
d'Astrophysique de Paris for
its hospitality and support, and to NASA for support under grants
NAG5-9266, NAG5-13288, and NAG5-10655.
JPD is supported by NSF grant AST-0070979 and
PHY-0205155, and NASA grant NAG5-9266.

\bibliographystyle{apj}
\bibliography{author}

\clearpage

\clearpage

\clearpage

\clearpage

\begin{table}
\begin{center}
\begin{tabular}{@{}lcccccccc}
\hline\hline
Model & $M_c/M_d$ & H/2 & $\langle \beta \rangle$ & resolution \\
\hline\hline
T & 0.5 & 0.6 & - & $128 \times 128$ \\
\hline
A1 & 0.5 & 0.6 & 400 & $256 \times 256$ \\
A2a & 0.5 & 0.6 & 1500 & $128 \times 128$ \\
A2b & 0.5 & 0.6 & 1500 & $256 \times 256$ \\
\hline
B & $\infty$ & 0.5 & 200 & $256 \times 256$ \\
\hline\hline
\end{tabular}
\end{center}
\caption{Model parameters.  Column $2$ gives the ratio between the
central mass $M_c$ and the disc mass $M_d$.  Column $3$ gives the height
of the computational domain (extending between $-H/2$ and $H/2$). Column
$4$ gives the ratio of the volume averaged initial values of the thermal
and
magnetic pressure.  Column $5$ gives the resolution of the run.}
\label{models}
\end{table}

\clearpage

\clearpage

\clearpage

\clearpage

\end{document}